\documentstyle[12pt,aasms4]{article}
\newcommand{\postscript}[2]
 {\setlength{\epsfxsize}{#2\hsize}
  \centerline{\epsfbox{#1}}}
\def\tempest%
{\begin{array}{ccc}
1 & 1 & 1 \\
1 & 1 & 1 \\
4 & 3 & 8
\end{array}}

\begin{document}

\title{Observational Evidence for the Effect of Amplifications\\
       Bias in Gravitational Microlensing Experiments}
\bigskip
\bigskip

\author{Cheongho Han}
\author{Youngjin Jeong}
\bigskip
\affil{Department of Astronomy \& Space Science, \\
       Chungbuk National University, Cheongju, Korea 361-763 \\
       cheongho@astro-3.chungbuk.ac.kr,\\
       jeongyj@astro-3.chungbuk.ac.kr}	
\authoremail{cheongho@astro-3.chungbuk.ac.kr}
\authoremail{jeongyj@astro-3.chungbuk.ac.kr}
\bigskip

\begin{abstract}
Recently Alard\markcite{alard1996} proposed to detect the shift 
of a star's image centroid, $\delta x$, as a method to identify the lensed
source among blended stars.  Goldberg \& Wo\'zniak\markcite{goldberg1997} 
actually applied this method to the OGLE-1 database and found that 7 out of 
15 events showed significant centroid shifts of $\delta x \gtrsim 0.2$ arcsec.
The amount of centroid shift has been estimated theoretically by 
Goldberg.\markcite{goldberg1997}  However, he treated the problem in general 
and did not apply it to a particular survey or field, and thus based his 
estimates on simple toy model luminosity functions (i.e., power laws). 
In this paper, we construct the expected distribution of $\delta x$
for Galactic bulge events by using the precise stellar LF observed by 
Holtzman et al.\markcite{holtzman1998} using HST.
Their LF is complete up to $M_I\sim 9.0$ ($M_V\sim 12$), corresponding
to faint M-type stars.  In our analysis we find that regular blending cannot 
produce a large fraction of events with measurable centroid shifts.
By contrast, a significant fraction of events would have measurable
centroid shifts if they are affected by amplification-bias blending.
Therefore, Goldberg \& Wo\'zniak's measurements of large centroid shifts
for a large fraction of microlensing events confirms the prediction of Han 
and Alard that a large fraction of Galactic bulge events are affected 
by amplification-bias blending.
\end{abstract}

\vskip50mm
\keywords{gravitational lensing --- luminosity function --- photometry}

\centerline{submitted to {\it The Astrophysical Journal}: Apr ??, 1998}
\centerline{Preprint: CNU-A\&SS-03/98}
\clearpage

\section{Introduction}

Experiments to detect Massive Astronomical Compact Halo Objects
(MACHOs) by monitoring the light variations of stars undergoing gravitational 
microlensing events are being carried out by 
several groups (MACHO, Alcock et al.\ 1997a; EROS, Ansari et al.\ 1996; 
OGLE, Udalski et al.\ 1997; DUO, Alard \& Guibert \ 
1997).\markcite{alcock1997a, ansari1996, udalski1997, alard1997b}
Since the lensing probability for a single source star is very low,
these searches are being conducted toward very dense star fields such as 
the Large Magellanic Cloud and Galactic bulge.
While searches towards these crowded fields result in an increased event rate,
it also implies that many of the observed light curves include light from
unresolved stars that are not being lensed: the blending problem.

Depending on the source for the blended light, blending can be classified 
into several types.
The first type, ``regular blending'', occurs when a bright source star
registered on the template plate is lensed and its flux is blended 
with the light from numerous faint unresolved stars below the 
detection limit imposed by crowding. 
Regular blending affects the results of lensing experiments in various ways.
First, it makes the measured event timescale shorter than the true one.
Since the lens mass scales as $M \propto t_{\rm E}^{2}$, the lens mass 
determined from the measured timescales will be underestimated and the 
lens population will be misinterpreted. 
In addition, since the optical depth is directly proportional to the 
summation of event timescales, i.e., $\tau \propto \sum_i t_{{\rm E},i}$, 
the Galactic MACHO fraction determined from the optical depth without a proper 
blending correction is subject to great uncertainty 
(Di Stefano \& Esin 1995).\markcite{distefano1995}

The second type of blending occurs if the source for the blended 
light is the lens itself: ``lens blending'' (Kamionkowski 1995; Buchalter, 
Kamionkowski, \& Rich 1996; Buchalter \& Kamionkowski 1997; Alard 
1997).\markcite{kamionkowski1995, bucjalter1996, buchalter1997, alrad1997}
Besides the effects of regular blending, lens blending has several
additional effects on the result of lensing experiments.
First, because detecting events due to lenses close to the observer is 
comparatively more difficult than detecting events produced by lenses 
near the source, lens blending makes the optical depth depend on the geometry 
of the lens system.
As a result, the matter distribution derived from the optical depth 
distribution deviates from its true one.
Secondly, lens blending causes the lensing optical depth to depend on
the lens mass function since more massive stars, which contribute more
to the total optical depth, tend to be brighter, resulting in a
larger blending effect (Nemiroff 1997; Han 1998).\markcite{nemiroff1997, 
han1998}

Finally, ``amplification bias'' blending occurs when one of several faint 
stars in the seeing disk below the detection limit is lensed, and the flux 
of the lensed star is associated with the flux from other stars in the 
integrated seeing disk (Nemiroff 1997).\markcite{nemiroff997}
In current experiments, photometry is carried out by comparing template 
images, in which only very bright stars are resolved and registered, with
a series of images taken of the same field.
The result is that in amplification-biased events the brightest star
appears to be the source because the lensed star is too faint to be resolved.
To be detected, the amplification-biased event must be highly amplified
to overcome the high threshold flux from the brighter star. 
Therefore, the mean detection probability to detect events for each source 
star will be very low.
However, if these faint stars comprise a significant fraction of the total 
number of stars, a considerable fraction of events might be amplification-biased 
(Han 1997b; Alard 1997a).\markcite{han1997b, alard1997a}
Moreover, due to the large amount of blended light from a bright star, 
the effects of blending for these events would be much more significant 
than those caused by other blending types.

There have been several methods proposed to correct for the blending problem.
The first method is to introduce an additional lensing parameter representing 
the residual flux from unresolved faint stars into the light curve fitting 
process.
However, this method suffers from large uncertainties in the derived
lensing parameters as a result of parameter degeneracies
(Wo\'zniak \& Paczy\'nski 1997).\markcite{wozniak1997}
Early-warning systems (for MACHO: Pratt et al.\ 1996;
for OGLE: Udalski et al.\ 1994; for PLANET: 
Albrow et al.\ 1995)\markcite{pratt1996, udalski1994, albrow1995}
allow one to construct lensing light curves with high time resolution 
and small photometric errors, enabling one to detect small color shifts caused by 
blending (Buchalter et al.\ 1996)\markcite{buchalter996}.
However, due to the narrow distribution of colors for Galactic bulge stars, 
the expected color shifts are small.
Han (1997b)\markcite{han1997} proposed to use the Hubble Space Telescope (HST)
to provide blending corrections.
By using the high resolving power of HST combined with color information
from ground-based observations, one can identify the lensed 
source star in the blended seeing disk, thus the uncertainty in
the derived timescale can be significantly reduced.
However, this method requires costly HST times.
One can also correct the blending effect statistically if the luminosity 
function(LF) of stars well below the detection limit can be constructed 
(Alcock et al.\ 1997b)\markcite{alcock1996}, but in this case we lose 
information about individual events.

Recently Alard (1996)\markcite{alard1996} proposed to detect the shift 
of a star's image centroid, $\delta x$, as a method to identify the lensed
source among blended stars.
Goldberg \& Wo\'zniak (1997)\markcite{goldberg1997}
actually applied this method to the OGLE-1 database and found that 7 out of
15 events showed significant centroid shifts of $\delta x \gtrsim 0.2$
arcsec.
The amount of centroid shift
has been estimated theoretically by Goldberg (1997).\markcite{goldberg1997}
However, he treated the problem in general and did not apply it
to a particular survey or field, and thus based his estimates
on simple toy model luminosity functions (i.e., power laws). 
In this paper, we construct the expected distribution of $\delta x$
for Galactic bulge events by using the precise stellar 
LF observed by Holtzman et al.\ (1998)\markcite{holtzman1998} using HST.
Their LF is complete up to $M_I\sim 9.0$ ($M_V\sim 12$), corresponding
to faint M-type stars. 
In our analysis we find that regular blending cannot produce
a large fraction of events with measurable centroid shifts.
By contrast, a significant fraction of events would have measurable
centroid shifts if they are affected by amplification-bias blending.
Therefore, Goldberg \& Wo\'zniak's measurements of large centroid shifts
for a large fraction of microlensing events confirms the prediction of Han (1997a) 
and Alard (1997) that a large fraction of Galactic bulge events are affected 
by amplification-bias blending.

\section{Centroid Shift}
Consider a seeing disk in which many closely-spaced stars are located with 
positions and fluxes given by ${\bf x}_i$ and $F_{0,i}$ respectively.
If one of these stars is gravitationally lensed and its flux is 
increased by $A_{\rm abs}$, the position of the center of light 
(CL) will shift toward the lensed star by an amount
$$
\delta x = \eta \left\vert\langle {\bf x}\rangle - 
{\bf x}_j\right\vert ;\qquad
\eta = {f(A_{\rm abs}-1) \over f(A_{\rm abs}-1)+1} =
       {A_{\rm obs}-1\over A_{\rm obs}},
\eqno(1)
$$
where $A_{\rm obs}=1+\kappa (A_{\rm abs}-1)$ is the observed amplification,
$\kappa=F_{0,j}/\sum_i F_{0,i}$ is the fractional light of the lensed 
source star before amplification,
and $\langle {\bf x}\rangle = (\sum_{i\neq j} F_{0,i} {\bf x}_i + 
F_{0,j} {\bf x}_j)/ \sum_{i} F_{0,i}$ is the position of the CL before 
amplification (Goldberg 1997).\markcite{goldberg1997}
Here the subscript ``$j$'' is used to designate the lensed star.

In Figure 1, we illustrate the centroid shifts for a simple case 
in which two stars are blended and the light fraction of the lensed source 
star is $\kappa=0.41$.
In the top panel, we present the point 
spread function (PSF) of each star.
Since the separation between the two stars is small, the integrated 
PSF appears to be a single star (left middle panel) with its center
at the CL (marked by a plus sign). 
In the rest of panels, we present the integrated PSFs and the position of  
CLs when $A_{\rm abs}=3$, 5, and 10 with respect to the original position 
of the CL when $A_{\rm abs}=1$.
As the amplification increases, one still cannot resolve individual PSFs, 
but the CL shifts increasingly towards the position of the lensed star.

\section{Expected Distribution of Centroid Shifts}

In this section we actually estimate the distribution of centroid shifts, 
$f(\delta x)$, which is expected for the microlensing experiment toward the 
Galactic bulge. 
Under current experimental strategy toward very crowded star fields, it is 
very likely that every event is affected by one of the three types of blending. 
However, the fraction of events affected by each type of blending is poorly known.
Among the three types of blending, Han (1998) found that although a 
significant fraction of lenses might be stars, most of them 
are very faint, and thus lens blending has only a small effect on the 
result of microlensing experiments. 
We begin by estimating the individual distributions of centroid 
shifts caused by regular and amplification-bias blending,
$f_{\rm reg}(\delta x)$ and $f_{\rm amp}(\delta x)$, and then estimate 
$f(\delta x)$ for various relative combinations of both types of blending.

To estimate $f(\delta x)$ it is essential to know how densely 
stars are crowded and what their brightness distribution is, i.e., the precisely 
determined LF of Galactic bulge stars.
We construct the LF by 
adopting the recent determination by Holtzman et al.\ (1998) using HST. 
Their LF is complete up to $M_I \sim 9.0$ ($M_V\sim12$), corresponding
to faint M-type stars. 
Although the expected contribution to the total event rate by stars
fainter than this limit would be very small, we extend the LF using 
the solar neighborhood LF from Gould, Bahcall, \& Flynn (1997) as
derived by HST imaging.
We note that the faint part of both the Galactic bulge and disk LFs
have similar shapes as seen by the good agreement between 
the two LFs in their overlapping 
region: $6.7 \lesssim M_I \lesssim 9.0$.
In Figure 2, we present the model LF.

To simulate observations of microlensing events, we begin by using a 
Monte Carlo method to produce stars whose fluxes are distributed according 
to the model LF.
At the same time, we assign positions to individual stars which are uniformly 
distributed throughout the seeing disk whose center lies at the CL.
The average seeing in current microlensing experiments is $\sim 2''$.
However, when a star is located at the edge of the seeing disk, 
one can isolate its amplified image from that of the integrated seeing disk.
We therefore set the size of the ``effective seeing disk'', the maximum 
unresolvable separation between stars, to $\theta_{\rm eff}=1''\hskip-2pt .5$. 
According to the model LF, there are on average $\sim 10$ stars within the 
effective seeing disk of angular area of 
$\pi (\theta_{\rm eff}/2)^2$. 
If the total flux of all stars within the effective seeing disk is greater 
than the detection limit imposed by crowding, we register this total as the 
template star brightness.
Current observations reach detection limit when the total density of 
stars toward Baade's Window is $\sim 10^6\ {\rm stars}/{\rm deg}^2$.
According to the model LF, this corresponds to $M_I \sim 3.1$ mag.

Next we select a lensed star.  
In the case of regular blending, the lensed star is the brightest 
star in the effective seeing disk. 
On the other hand, in the case of amplification-biased events, the lensed
star is one of the relatively faint stars in the effective seeing disk.
Once the source stars are chosen, we produce test events whose lensing impact 
parameters, $\beta$, are distributed randomly in the range from 0 to 1. 
Among these events, detectable events should satisfy the condition that
the integrated flux at peak amplification is greater than the
flux before amplification by a factor of 1.34, the threshold amplification,
i.e.,   
$$
{\sum_{i\neq j} F_{0,i} + A_{\rm max}F_{0,j}\over 
\sum_i F_{0,i}} \geq 1.34,
\eqno(2)
$$
where $A_{\rm max} = (\beta^2 +2)/\beta (\beta^2+4)^{1/2}$.
Finally, we compute the centroid shifts of individual detected events 
using equation (1).

In the upper panel of Figure 3, we present the expected distribution 
of centroid shifts caused by regular (dotted line) 
and amplification-bias blending (dashed line).
In the lower panel, we present the fraction of events with centroid 
shifts greater than a certain limiting value $\delta x_{\rm lim}$, i.e.,
$1-\int_0^{\delta x_{\rm lim}} f(\delta x) d\delta x$.
In both panels, the distributions are normalized for 
a single source star, i.e., $\int_0^\infty f_{\rm reg}(\delta x) d\delta x 
=\int_0^\infty f_{\rm amp}(\delta x) d\delta x = 1.0$
and the range $0\leq \delta x \leq 1$ is divided into 50 intervals.
One finds several differences between the two distributions.
First, while the peak of $f_{\rm reg}(\delta x)$ occurs at $\delta x = 0$ 
and the distribution decreases rapidly with increasing values of $\delta x$, 
the distribution $f_{\rm amp}(\delta x)$ peaks at $\delta x\sim 0\farcs2$
and decreases gradually toward both smaller and larger values of $\delta x$.
In addition, while a significant fraction of amplification-biased events,
$\sim 50\%$, has centroid shifts greater than $\delta x_{\rm lim}=0.2$ 
arcsec, the fraction of regular blended events with 
$\delta x \ge 0.2$ arcsec is just $\sim 2\%$.

Once the distribution of centroid shifts caused by each type of blending is 
obtained, the actual centroid shift distribution is 
obtained by 
$$
f(\delta x) = (1-{\cal F}_{\rm amp}) f_{\rm reg}(\delta x)
              + {\cal F}_{\rm amp} f_{\rm amp}(\delta x)
\eqno(3)
$$
with the known amplification-biased event fraction of ${\cal F}_{\rm amp}$.
Han (1997a) estimated that ${\cal F}_{\rm amp}\sim 40\%$ of events 
detected toward the Galactic bulge are affected by amplification-bias blending.
However, to see the dependency of $f(\delta x)$ on the value of
${\cal F}_{\rm amp}$, we examine various amplification-biased event fractions,
${\cal F}_{\rm amp}= 20\%$, 40\%, and 80\%.  The resulting 
distributions of $f(\delta x)$ and 
$1-\int_0^{\delta x_{\rm lim}} f(\delta x) d\delta x$ are presented in 
Figures 4 and 5, respectively.
From these figures, one first finds that the distribution $f(\delta x)$ 
changes significantly for different values of ${\cal F}_{\rm amp}$. 
Secondly, in order to produce large centroid shifts for a significant
number of events, a large fraction of Galactic bulge events must
be affected by amplification-bias blending. 
Therefore, the recent finding by Goldberg \& Wo\'zniak (1997) that a
considerable fraction of Galactic bulge events exhibit considerable centroid
shifts confirms the predictions of Han (1997b) and
Alard (1997a)\markcite{han1997, alard1997a} that a substantial fraction of 
Galactic bulge events are affected by amplification-bias blending.

\section{Discussion}
The fact that events affected by amplification-bias blending produce
large $\delta x$ while the centroid shifts caused by
regular blending are small can be understood analytically in the 
following way.
To produce centroid shifts large enough to be measured, events should
satisfy two conditions.
First, the event should be highly amplified. 
For a very low amplification event ($A_{\rm abs}\sim 1$), 
the expected centroid shift will be small since $A_{\rm abs}-1 \sim 0$
and thus $\eta \sim 0$ in equation (1).  On the other hand,
for a very high-amplification event ($A_{\rm abs}\sim \infty$), one finds
$\delta x\sim \left\vert\langle {\bf x}\rangle -{\bf x}_{j}\right\vert$ 
since the factor $\eta$ approaches unity.
However, not all events that satisfy the first condition produce large 
centroid shifts.
The second condition is
that the lensed star should be one of the faint
stars in the blended seeing disk.
If the lensed star is the brightest one in the effective seeing disk and
its flux dominates the flux over those from other faint blended stars, i.e.,  
$\kappa\sim 1$, the position of the CL before gravitational amplification will be 
very close to that of the lensed star, resulting in a small amount of shift
($\left\vert\langle {\bf x}\rangle-{\bf x}_j\right\vert \sim 0$) since
$\sum_{i\neq j} F_{0,i}{\bf x}_i + F_{0,j}{\bf x}_j\sim F_{0,j}{\bf x}_j$ and
$\sum_i F_{0,i} \sim F_{0,j}$.
On the other hand, if the lensed source is very faint, i.e., $\kappa\sim 0$, 
the light from the source star has negligible effect on the position of CL,
resulting in high possibility of a large centroid shift.  In summary,
the expected centroid shifts for various extreme cases of
amplification and source star brightness are:
$$
\cases{
\eta\sim 0\ {\rm and}\ \delta x\sim 0   & for a low amplification event \cr
                                        & ($A_{\rm abs}\sim 1$) \cr
\delta x\sim \left\vert\langle{\bf x}\rangle-{\bf x}_j\right\vert\sim 0 
                                        & for a high-amplification event\cr
	                                & with luminous source \cr
				        & ($A_{\rm abs}\sim \infty$, 
                                          $\kappa\sim 1$) \cr
\delta x\sim \left\vert\langle{\bf x}\rangle-{\bf x}_j\right\vert       
                                        & for a high-amplification event\cr
                                        & with faint source \cr
                                        & ($A_{\rm abs}\sim \infty$, 
			                  $\kappa\sim 0$). \cr
}
\eqno(4)
$$

For amplification-biased events, source stars are in general
very faint ($\kappa\sim 0$), mostly far below the detection limit.
Despite their low luminosities, the fact that they are detected implies that
the source stars are highly amplified ($A_{\rm abs}\sim \infty$). 
Therefore, the conditions for large centroid shifts 
agree well with those for amplification-biased events.
On the other hand, regular blended events do not meet these conditions.
First, due to relatively small amount of blended light, regular blended 
events do not need to be highly amplified for detection ($A_{\rm abs}\sim 1$).
Although they can be highly amplified ($A_{\rm abs}\sim \infty$), 
the dominance of their fluxes ($\kappa\sim 1$) over those from other faint 
blended stars will result in small $\delta x$.

As demonstrated by the large centroid shifts for a significant fraction of
events, the effect of amplification-bias blending on the results of 
lensing experiments is important.
However, the methods to correct for the blending effect mentioned in \S\ 1 
have various limitations in application.
One very simple but very practical method to minimize the effect of 
amplification-bias blending is to monitor only very bright stars.
With increasing reference image brightness, the required amplification 
for detection becomes higher, resulting in a lower probability
of amplification-biased events. 
To show how the blending effect decreases with increasing reference 
image brightness, we simulate Galactic bulge events which are expected to be
detected for various threshold reference image brightnesses.
For each event, we compute the light fraction of the source star $\kappa$
and the timescale decrease factor $\eta$.  The distributions of $\kappa$ and
$\eta$ are shown in the upper panel of Figure 6.
For a given fraction of source star flux, the observed timescale is reduced by
$\eta=t_{\rm eff}/t_{\rm E}=
\left[ 2(1-A_{\rm min}^{-2})^{-1/2}-2 \right]^{1/2};
\ A_{\rm min} = 0.34/\kappa + 1$.
Here events are assumed to be detected 
as long as they can amplify the reference image flux by more than
a factor of 1.34. 
However, highly blended events will have short $t_{\rm eff}$, resulting 
in low detectability. 
Therefore, we correct the distributions by the detection efficiency.  
We assume that the detection efficiency is linearly proportional to 
the timescale decrease factor, i.e., $\epsilon \propto \eta$.
The efficiency-corrected distributions $f(\kappa)$ and $f(\eta)$ are 
presented in the middle panels.
In the lower panels we present the distributions of 
the fraction of events with $\kappa \ge \kappa_{\rm lim}$,
$1-\int_{0}^{\kappa_{\rm lim}} f(\kappa)d\kappa$, and 
$\eta \ge \eta_{\rm lim}$, $1-\int_{0}^{\eta} f(\eta')d\eta'$.
From these distributions one finds that under the current threshold reference 
image brightness of $M_I\sim 3$, the fraction of events
with little blending effects ($\kappa \gtrsim 0.9$ or $\eta \gtrsim 0.9$)
is $\lesssim 10\%$.
However, as the brightness of the threshold reference image increases, this
fraction gradually increases until it becomes $\sim 80\%$ when
only stars brighter than $M_I\sim 0$ are monitored, which corresponds to
the brightness of Galactic bulge clump giant stars 
(Paczy\'nski \& Stanek 1998).\markcite{paczynski1998} 
The MACHO group (1997a)\markcite{alcock1997a} already applied this method 
and their optical depth determination is based on clump giant stars.
However, by monitoring significantly fewer stars at a decreased event rate,
the statistical precision of the lensing experiments will be lowered.
A more general solution for the blending correction is
provided by the rapidly progressing image subtraction technique
(Alard \& Lupton 1997b; Tomaney 1998)\markcite{alard1997c, tomaney1998}
which is also being applied to detect microlensing events towards M31 
by the Colombia-Vatt group (Crotts \& Tomaney 1996; 
Tomaney \& Crotts 1996).\markcite{crotts1996, tomaney1996}

\acknowledgements
We would like to thank M.\ Everett and S.\ Gaudi for making useful comments.
This work was supported by a grant 981-0203-010-1 from Korea Science
and Engineering Foundation.

\clearpage

\postscript{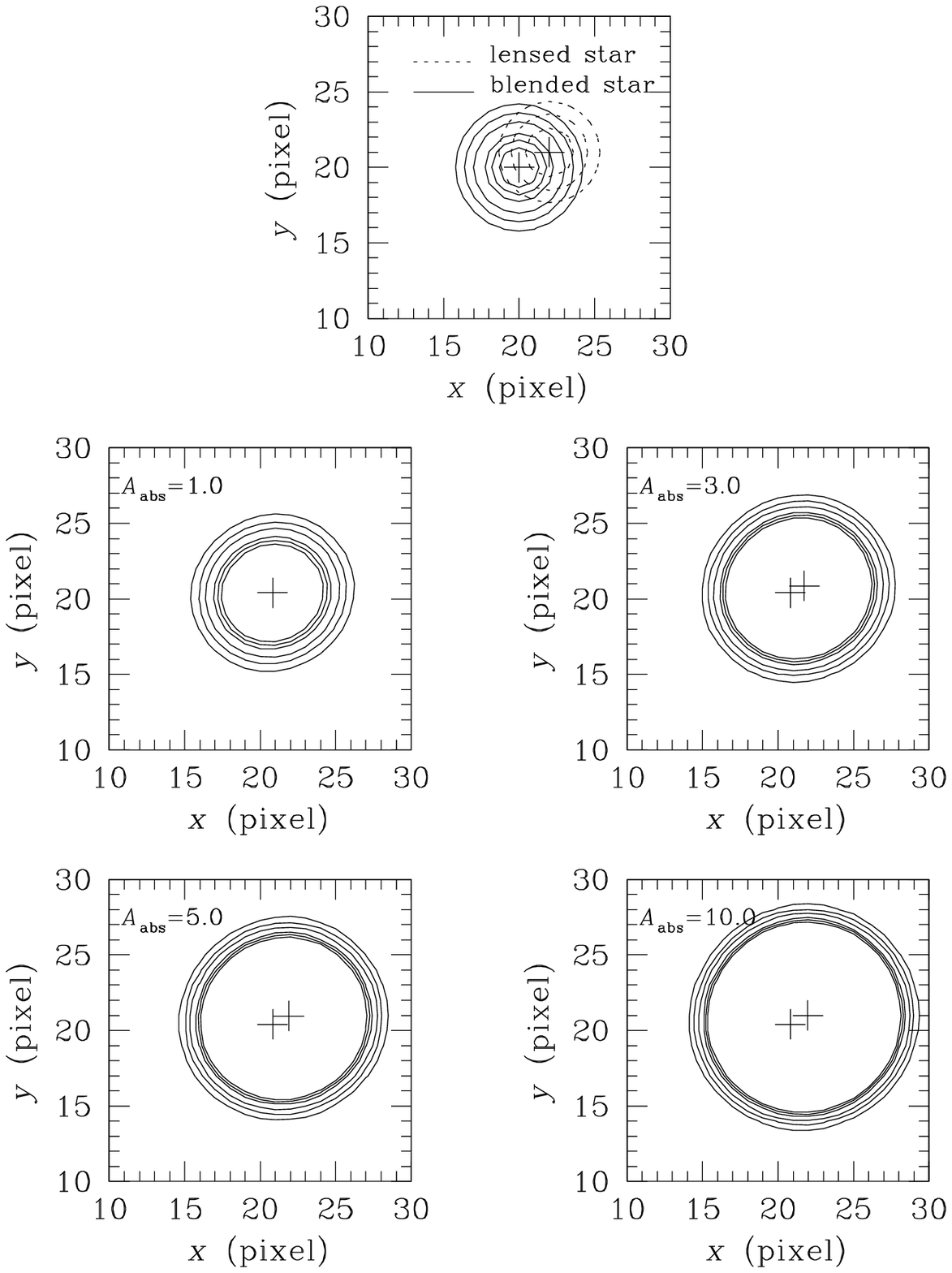}{0.95}
\noindent
{\footnotesize {\bf Figure 1:}\
Illustration of the centroid shifts for a simple case in which two stars 
are blended and the fraction of light from the lensed source star is 
$\kappa=0.41$.
In the top panel of the figure, we present the PSF of each star.
Since the separation between the two stars is small, the integrated
PSF appears to be a single star (left middle panel) with its center
at the CL (marked by a plus sign).
In the rest of panels, we present the integrated PSFs and the
position of CL for $A_{\rm abs}=3$, 5, and 10 with respect to the 
original position of the CL before amplification.
As the amplification increases, one still cannot resolve individual PSFs,
but the CL shifts increasingly towards the position of the lensed star.
}

\postscript{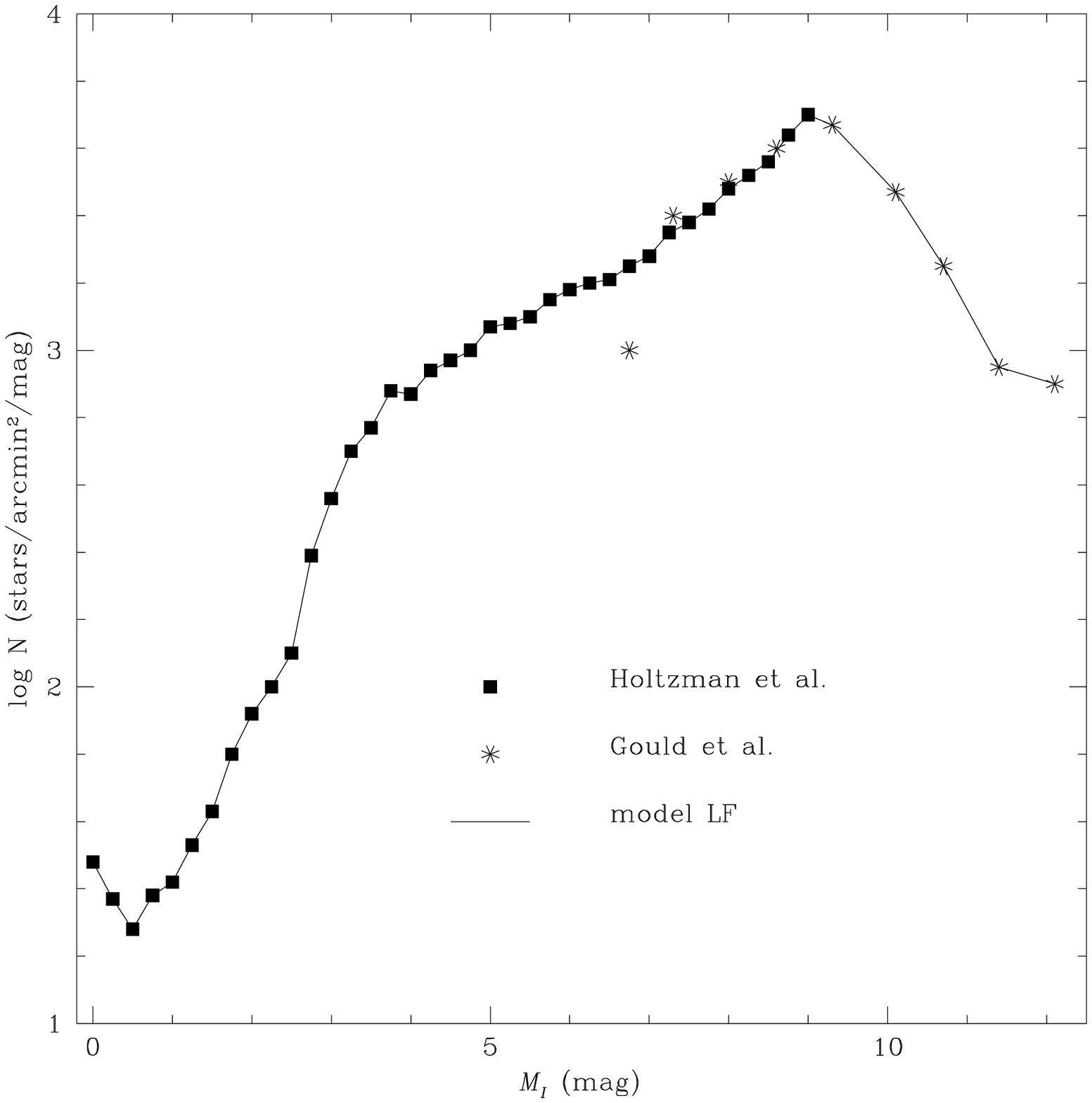}{0.95}
\noindent
{\footnotesize {\bf Figure 2:}\
The model LF.
We construct the LF up to  $M_I\sim 9$ by adopting the recent determination 
by Holtzman et al.\ (1998) using HST, marked by filled squares, and 
further extend the LF by using the solar neighborhood LF from 
Gould, Bahcall, \& Flynn (1997), marked by asterisks.
We note that the faint part of the both Galactic bulge and disk LFs
have similar shapes as seen by the good agreement between
the two LFs in their overlapping region $6.7 \lesssim M_I \lesssim 9.0$.
}

\postscript{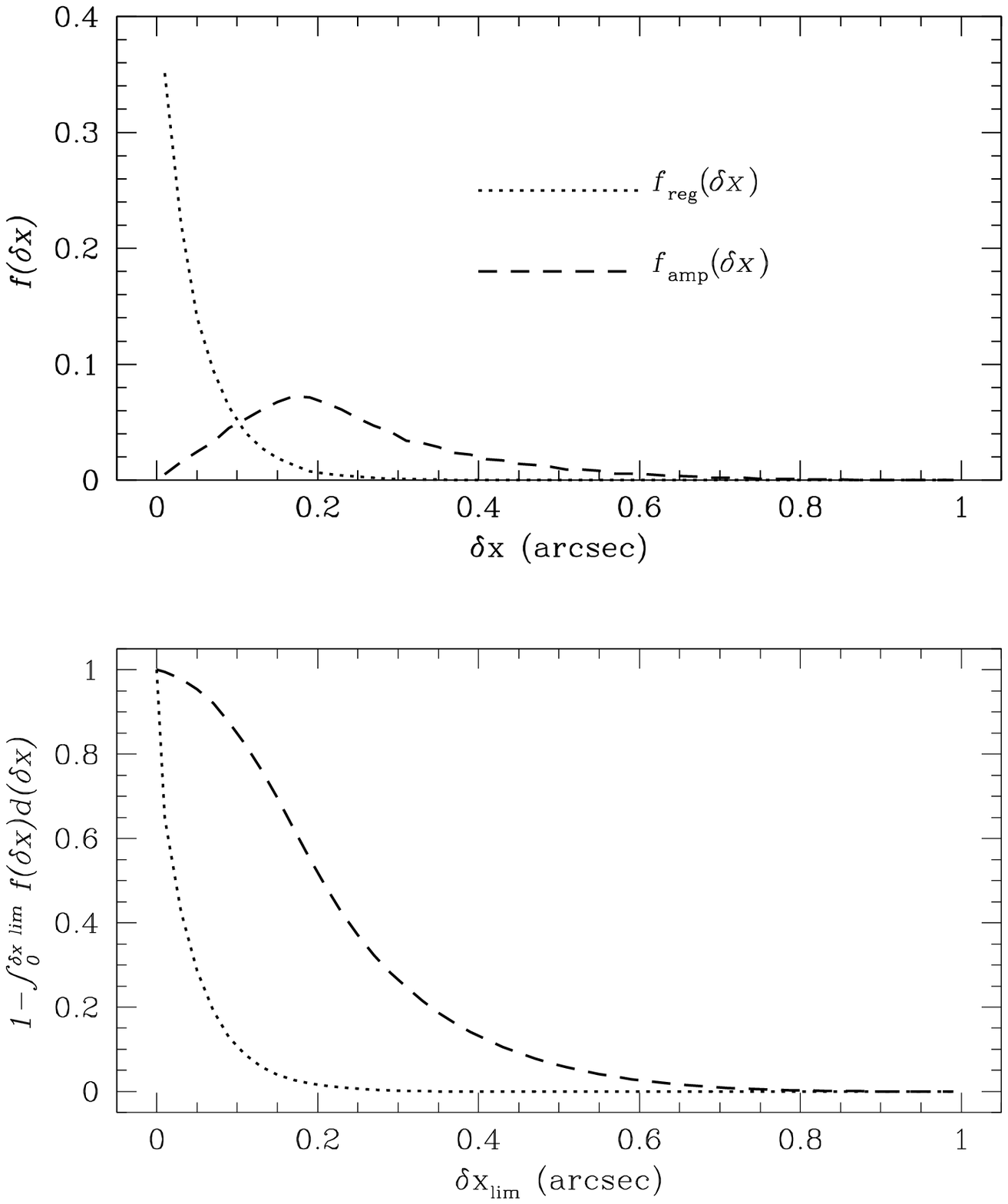}{0.95}
\noindent
{\footnotesize {\bf Figure 3:}\
Upper panel: The expected distributions
of centroid shifts caused by regular [$f_{\rm reg}(\delta x)$, dotted line]
and amplification-bias blending [$f_{\rm amp}(\delta x)$, dashed line].
Lower panel: the fraction of events with centroid
shifts greater than a certain limiting value $\delta x_{\rm lim}$, i.e.,
$1-\int_0^{\delta x_{\rm lim}} f(\delta x) d\delta x$.
In both panels, the distributions are normalized for a single source star, 
i.e., $\int_0^\infty f_{\rm reg}(\delta x) d\delta x
=\int_0^\infty f_{\rm amp}(\delta x) d\delta x = 1.0$, and the range
$0\leq \delta x\leq 1$ is divided into 50 intervals.
}

\postscript{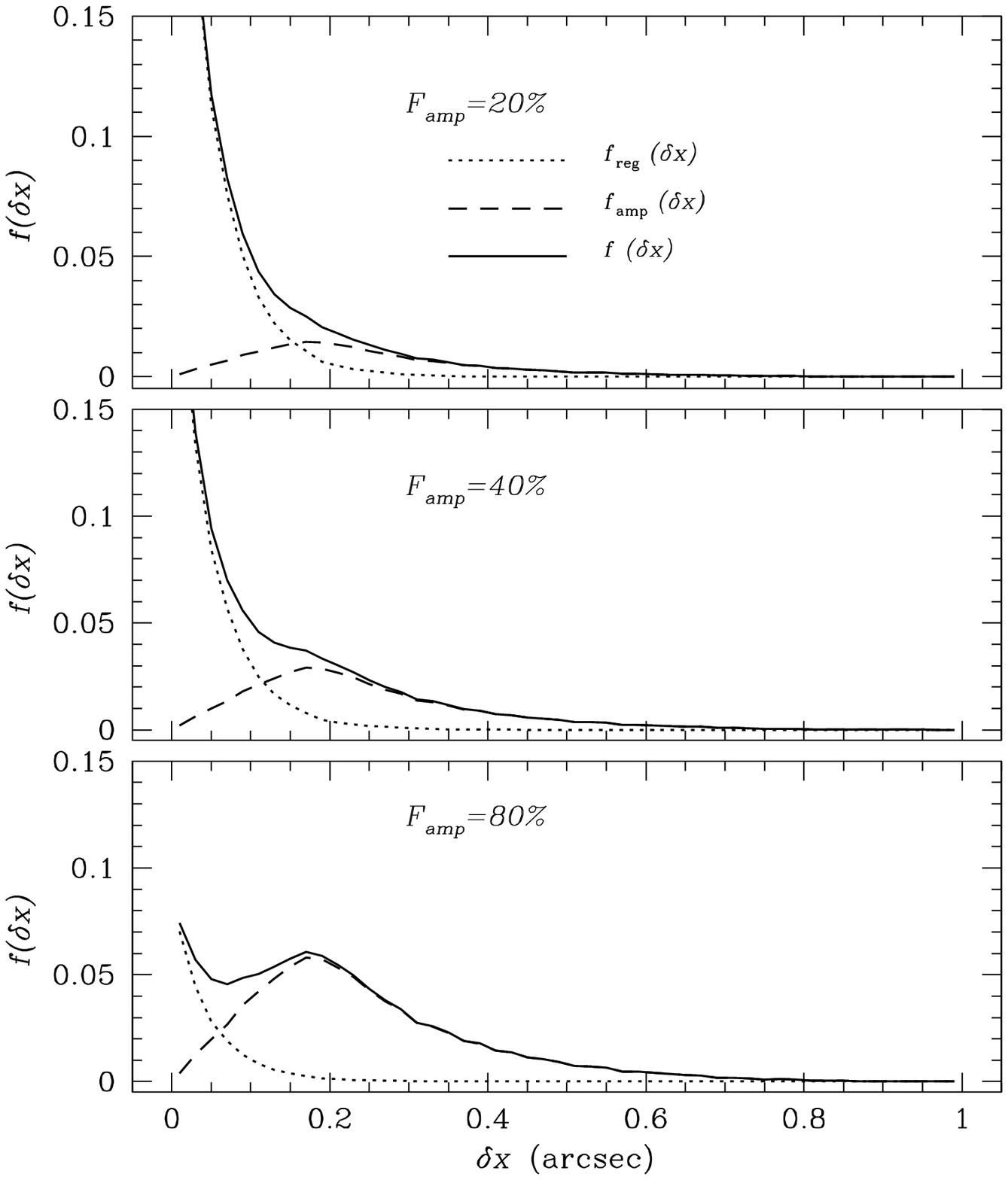}{0.95}
\noindent
{\footnotesize {\bf Figure 4:}\
The expected centroid shift distribution for events towards the
Galactic bulge for various amplification-biased event fractions 
of ${\cal F_{\rm amp}}=20\%$, 40\%, and 80\%.
From the known distributions of centroid shifts due to regular, 
$f[{\rm reg} (\delta x)]$, and amplification-bias blending,
$f[{\rm amp} (\delta x)]$, the actual centroid shift distribution is 
obtained by $ f(\delta x) = (1-{\cal F}_{\rm amp}) f_{\rm reg}(\delta x)
              + {\cal F}_{\rm amp} f_{\rm amp}(\delta x)$.
}

\postscript{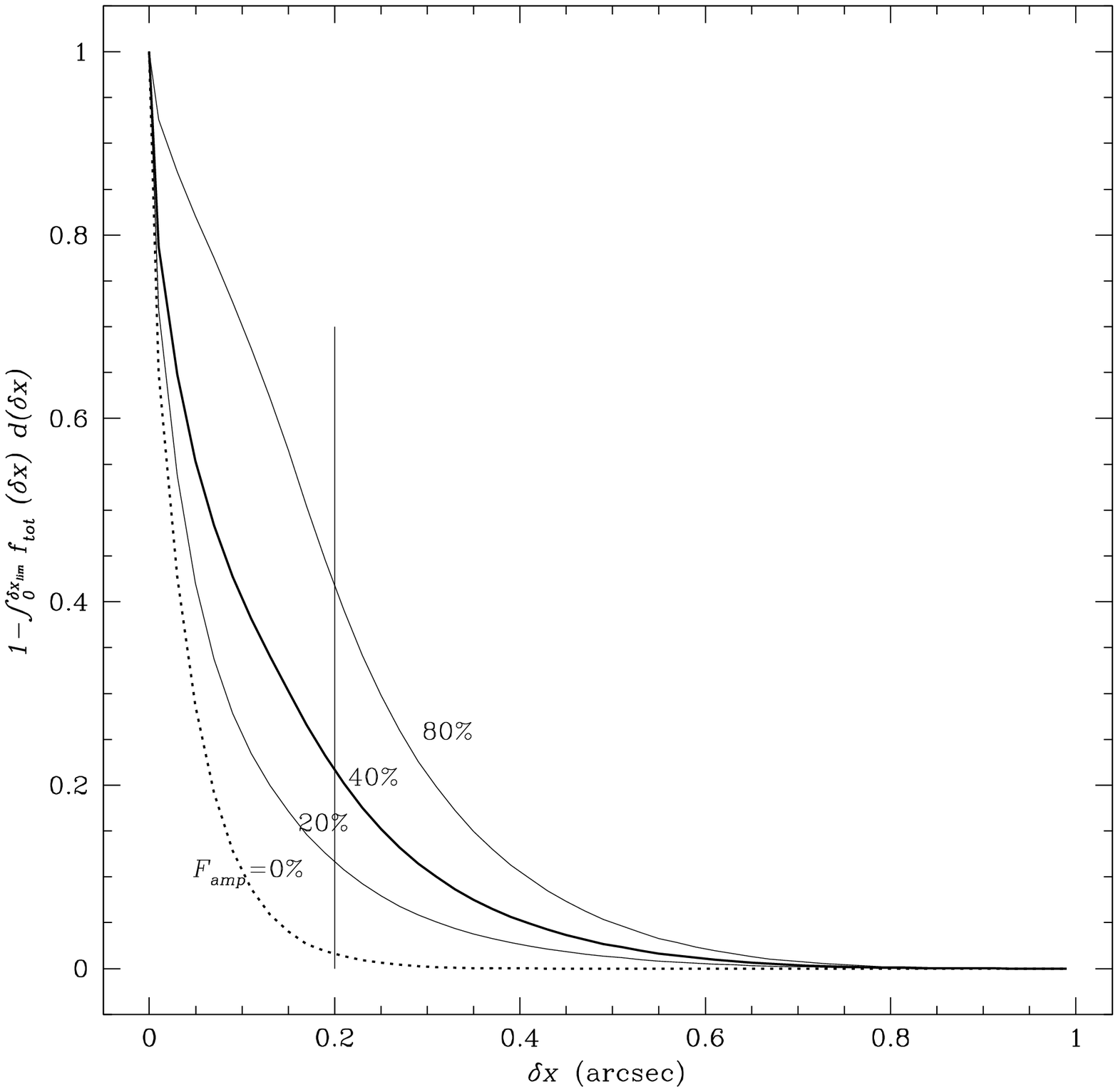}{0.95}
\noindent
{\footnotesize {\bf Figure 5:}\
The distribution of the fraction of events with centroid shifts 
greater than a limiting value of $\delta x_{\rm lim}$ for various 
amplification-biased event fractions of 
${\cal F}_{\rm amp}=0\%, 20\%$, $40\%$, and $80\%$.
If all events are free from amplification-bias blending 
(${\cal F}_{\rm amp} = 0$), only $\sim 2\%$ of events would have 
centroid shifts with $\delta x \ge 0.2$ arcsec.
On the other hand, to produce large centroid shifts for a significant
number of events, a large fraction of Galactic bulge events should 
be affected by amplification-bias blending. 
}

\postscript{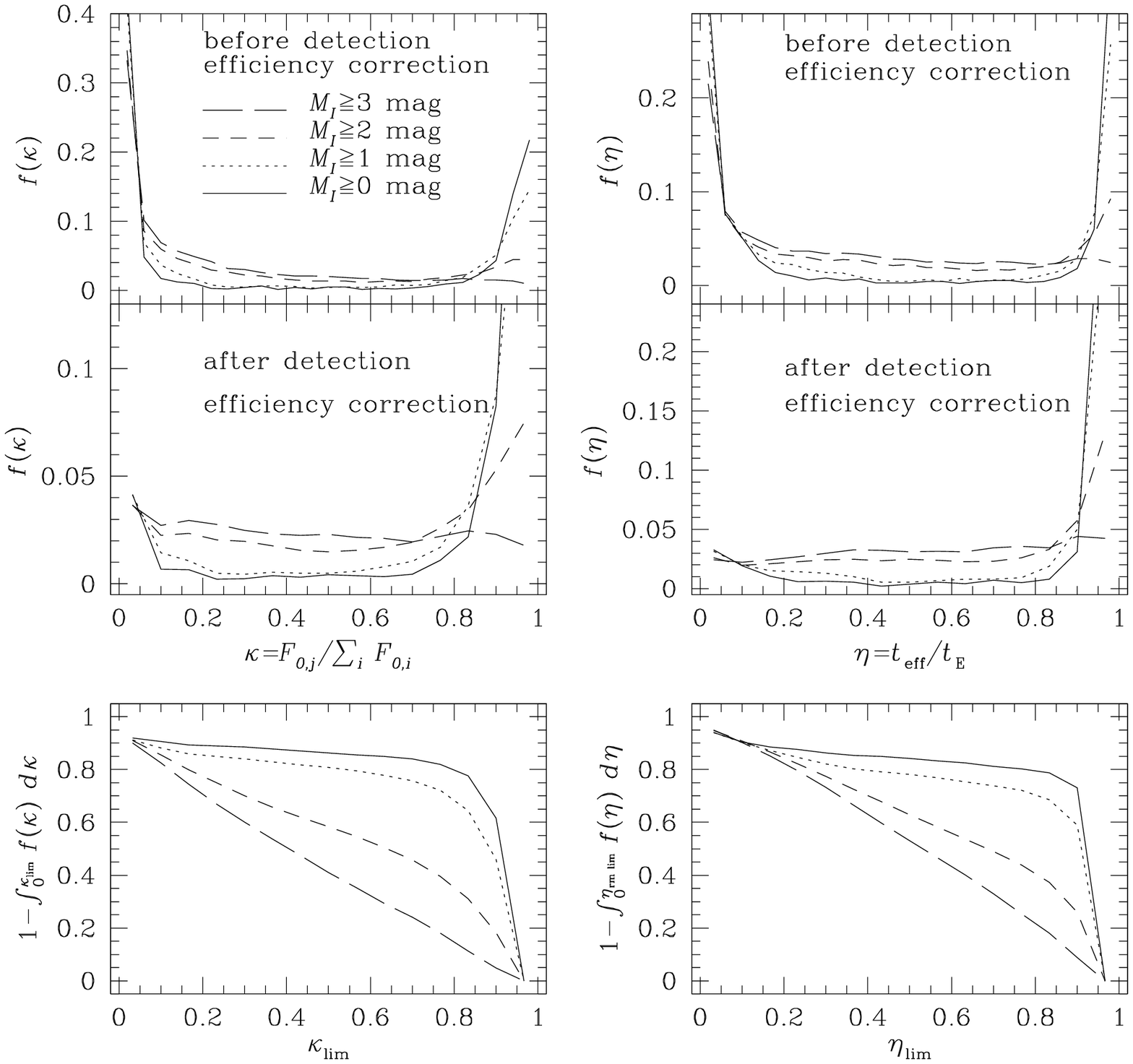}{0.95}
\noindent
{\footnotesize {\bf Figure 6:}\
Correction for amplification-bias blending by monitoring bright source stars.  
Upper panels: The expected distributions of the fractional light $\kappa$ 
and timescale decrease factor $\eta$ for various threshold brightness of 
reference image.
Middle panels: same as upper panels, but the distributions are corrected by
the detection efficiency.  We assume that the detection efficiency is 
linearly proportional to the timescale decrease factor, i.e., 
$\epsilon \propto \eta$.  
Lower panels: The distributions of events with the fractional light 
$\kappa \ge \kappa_{\rm lim}$ (left panel) and the timescale decrease factor
$\eta \ge \eta_{\rm lim}$ (right panel).
}

\end{document}